\documentclass[aps,preprint,showpacs,smsmath,amssymb]{revtex4}
\usepackage{epsfig}
\usepackage{graphicx}

\newcommand{\allgraphs}{{\mathcal G}}
\newcommand{\N}{{\rm I\!N}}

\newcommand{\pprime}{{\prime\prime}}

\newcommand{\be}{\begin{equation}}
\newcommand{\ee}{\end{equation}}
\newcommand{\bd}{\begin{displaymath}}
\newcommand{\ed}{\end{displaymath}}

\newcommand{\bra}{\langle}
\newcommand{\ket}{\rangle}

\newcommand{\order}{{\cal O}}

\newcommand{\bnull}{\mbox{\boldmath $0$}}
\newcommand{\bc}{\mbox{\boldmath $c$}}

\newcommand{\bk}{\mbox{\boldmath $k$}}

\newcommand{\bx}{\mbox{\boldmath $x$}}

\newcommand{\bomega}{\mbox{\boldmath $\omega$}}

\newcommand{\bxi}{\mbox{\boldmath $\xi$}}

\begin{document}

\title{Entropies of complex networks with hierarchically constrained topologies}

\author{Ginestra Bianconi$^1$, Anthony C. C. Coolen$^{2,3}$,
  Conrad J. Perez Vicente$^4$}

\affiliation{$^1$ Abdus Salam
International Center
for Theoretical Physics,
Strada Costiera 11,
34014 Trieste,
Italy\\
$^2$ Department of Mathematics, King's College London, The Strand,
London WC2R 2LS, United Kingdom\\
$^3$ Randall Division of Cell and Molecular Biophysics,
King's College London, New Hunt's House,
London SE1 1UL, United Kingdom\\
$^4$ ~Departament de F\'isica Fonamental, Facultat de F\'isica,
 Universitat de Barcelona, 08028 Barcelona, Spain}

\begin{abstract}

The entropy of a hierarchical network
topology in an ensemble of sparse random networks, with  "hidden
variables" associated to its nodes, is the log-likelihood
that a given network topology is
present in the  chosen ensemble. We obtain a
 general formula for  this entropy, which has a clear interpretation
 in  some simple limiting cases.
The results  provide new keys with which to solve the
general problem of ``fitting'' a given network with an appropriate
ensemble of random networks.
\end{abstract}

\maketitle
\section{Introduction}

The entropy is a key concept in information theory
\cite{information} and in the theory of dynamical systems \cite{ds}.
In information theory, the problem of inference of a probability
distribution on the basis of finite number of  independent observations
is usually addressed  using the maximum likelihood principle or via the minimization of the
Kullback-Leibler distance between the given (empirical) distribution and
the inferred one.
Recently, several studies have  extended the tools of
 information theory along these lines in order to measure
the performance of filtering procedures of correlation matrices  in the case of
multivariate data \cite{Mantegna,Biroli}.
In  the framework of graph theory the
large deviations of the ensemble of  random Erd\"os and Renyi graphs   where derived by studying  the free energies of
statistical mechanics models defined on them \cite{Hartmann,Rivoire}.
 There is now increased interest, in the community of  complex networks
\cite{AlbertBarabasi,Dorogovtsev,Latora}, in
the definition of entropy measures  that are related with  the networks' topological
structure \cite{europhys} or with diffusion processes defined on them \cite{Latora2}.
The inference problem applied to   complex networks can
be formulated as the
identification of  the  ensemble of networks which
retains the  essential structural characteristics and complexity of a
given real network realization.
The identification  of this ensemble,
is an active field of research.
One aims to fit a given specific network with a
suitable network ensemble that retains some information on its structure.
Newman has proposed this approach to find the community structure in a
given network \cite{Newman}.
Later, this method  has been extended    to define ensembles of  networks
 that have other topological
characteristics  in common
with the real network, such as the degree sequence and/or  the degree correlations.
As we add further features that
a desired ensemble is to have in common with a given real network, we effectively consider
ensembles with decreasing cardinality.
The cardinality of an ensemble of networks with a given topology has
attracted the  attention of  the graph theory community
\cite{Enumeration1,Enumeration2,Wormald}, and more recently also of the
statistical mechanics community \cite{europhys}.

In this paper we  evaluate  the entropy of a given hierarchical
topology in a "canonical" or "hidden variable" ensemble, i.e. we calculate the
normalized logarithm of the probability that a given topology appears
in this ensemble.
By hierarchical topology we will mean the set of the generalized degrees
of the nodes, defined as the sequence $\bk_i=(k_i^1,k_i^2,\ldots, k_i^L)$ of the
number of nodes at distance $1, 2,\ldots,L$  from the node $i$.
The "canonical'' or ``hidden variable'' \cite{hv1,hv2,hv3,hv4,Boguna}  ensembles are generalization of the $G(N,p)$ ensemble for
heterogeneous nodes. The hetereogeneity of the nodes is  described in terms of some "hidden variables''
$x_i$, defined on each node $i$ of the network, and
the probability $p_{ij}$ of a link between a node $i$ and a node $j$
is not $p$ as in $G(N,p)$ but it is a general function $Q(x_i,x_j)$ of the
 hidden variables at $i$ and $j$ nodes.
These ensembles  correspond to networks which satisfy soft constraints,
for example the degree of a node is not fixed, but  only  the average degree
of each node is fixed, allowing for  Poissonian fluctuations.

We derive a general formula for the  entropy of a given topology in a
"canonical" ensemble using ideas and methods from
the  study of diluted  combinatorial optimization
problems and statistical mechanical systems on sparse networks
\cite{Monasson,monasson-zecchina98,monasson-zecchina982,monasson-zecchina99,VCising1,ColoringRG1,Leone,wemmenhove-coolen03,perez-skantzos03,arbitrary_degrees,
SmallWorld,replTransMat,DRTfc, Goos,
Coolen_etal2005,Nikos2005,LLmodel,Mozeika,ColoringRG2,CDMA3,CDMA4}.
In the simple case where we study the likelihood of a degree
distribution of a network belonging to the chosen ensemble the
entropy is found to be the Kullback-Leibler distance
between the probability distribution of the
degrees and the expected probability of the typical
topology of the network.

The paper is structured as follows: in section II we introduce the
definition of the problem, in section III we provide  the asymptotic
entropy expression of the network topology in a given ensemble, in section
IV we study the form that the entropy takes in special and relevant
cases, and the conclusions are presented in  section V.

\section{Formulation of the problem and definitions}

To model the essential properties of a real network it is useful to think of it
as an instance of an ensemble of networks.
The ensemble can be either  "microcanonical" or  "canonical" depending
on whether the networks in the ensemble are subject to hard or soft constraints.
The main example of what we call a "microcanonical" ensemble is
$G(N,M)$ where the number of links is fixed to be exactly $M$, and the main
example of "canonical" ensemble is  $G(N,p)$ in which only the average
number of links $\bra{M}\ket=pN(N-1)/2$ is fixed.
These ensembles can be generalized to ensembles of random graphs
with a given degree sequence and with a given hidden variable distribution.
In this paper we will calculate the entropy of a given network
topology (defined in terms of  its hierarchical structure) in a
general "canonical'' ensemble.  This entropy is defined as the probability that
the given network topology is found in the "canonical'' network
ensemble under consideration.

\subsection{ ``Canonical''  ensembles}

We consider networks characterized by $N$ nodes (or `sites')
labeled $i=1,\ldots, N$, and a symmetric matrix $\bc$ with entries
$c_{ij}\in\{0,1\}$ that specify whether ($c_{ij}=1$) or not
($c_{ij}=0$) nodes $i$ and $j$ are connected. We choose $c_{ii}=0$
for all $i$. We write the set of all such undirected networks as
$\allgraphs=\{0,1\}^{\frac{1}{2}N(N-1)}$. On this set $\allgraphs$ we introduce the following
probability measure, in order to define an ensemble $\{\allgraphs,W\}$ of random networks:
\begin{eqnarray}
W(\bc|\bx)&=& \prod_{i<j}\Big[\frac{c}{N}Q(x_i,x_j)\delta_{c_{ij},1}+(1-\frac{c}{N}Q(x_i,x_j))\delta_{c_{ij},0}\Big]
\label{eq:ensemble}
\end{eqnarray}
The $\{x_i\}$ represent `hidden variables', drawn for each site independently with statistics $p(x)$ to be defined later,
and the function $Q(x,x^\prime)\geq 0$ is chosen such that $\sum_{xx^\prime}p(x)p(x^\prime)Q(x,x^\prime)=1$. The latter condition
ensures that asymptotically $c$ represents the average connectivity,
viz. $\lim_{N\to\infty}\bra N^{-1}\sum_{ij}c_{ij}\ket=c$.
Note that throughout    this paper the 'hidden variables' $\{x_i\}$ can
be scalar, discrete or multidimensional.

\subsection{Hierarchical constraints topologies}

Next we introduce  a hierarchy of single-site observables with the objective to characterize with increasing precision
the local topology of a network $\bc\in\allgraphs$. They can be interpreted as generalized degrees $\bk_i(\bc)=(k_i^1(\bc),\ldots,k_i^L(\bc))$ of
individual nodes $i$:
\begin{eqnarray}
k_i^\ell(\bc)&=& \sum_{j_1 \ldots j_\ell}c_{ij_1}c_{j_1 j_2}\ldots c_{j_{\ell-1} j_\ell}~\in~\{0,1,2,\ldots,N^\ell\}
\label{eq:generalized_degrees}
\end{eqnarray}
In the absence of local loops, $k_i^\ell(\bc)$ measures the size (measured in number of nodes) of the local environment of node $i$, at a distance of $\ell$ links. However, in this tree the nodes are counted with a multiplicity equal to their number of descendants encountered;
similarly, in the case of local loops, nodes that can be visited from site $i$ via multiple routes of length $\leq \ell$ are counted with this multiplicity. Note that
$k_i^1(\bc)=\sum_j c_{ij}$ is the ordinary degree of node $i$, and that
(\ref{eq:generalized_degrees}) can also be written as
\begin{eqnarray}
&&
k_i^1(\bc)=\sum_j c_{ij},~~~~~~k_i^{\ell+1}(\bc)=\sum_j c_{ij}k_j^{\ell}(\bc)
\label{eq:generalized_degrees2}
\end{eqnarray}
By definition, if $k_i^1(\bc)=0$ then $k_i^\ell(\bc)=0$ for all $\ell$.
 It is now natural to characterize the {\em global}
topology of a network
$\bc$ either by giving its $N$ generalized degree vectors $\{\bk_1(\bc),\ldots,\bk_N(\bc)\}$
themselves, or
by giving the collective generalized degree
statistics, conditioned on the values of the hidden variables, i.e.
\begin{eqnarray}
 && P(\bk|x,\bc)= P(k_1,\ldots,k_L|x,\bc)=\frac{1}{Np(x)}\sum_{i=1}^N
\delta_{\bk,\bk_i(\bc)}\delta(x-x_i)
 \label{eq:general_distribution}
\end{eqnarray}
We adopt the convention that always $\bk=(k_1,\ldots,k_L)\in\N^L$,
unless indicated otherwise.

\subsection{Entropy of a network contraint topology in a given ensemble }

Our goal is to quantify to what extent the above characterization of networks, by
the generalized degrees $\{\bk\}\equiv\{\bk_1,\ldots,\bk_N\}$ or by the degree statistics $P_{L}(\bk)$, specifies their micro-structure.
This can be measured by the effective number of networks in the ensemble $\{\allgraphs,W\}$ that meet the relevant contraints, i.e.
(apart from a constant) by the Boltzmann entropies:
\begin{eqnarray}
\hspace*{-2mm}
{\rm constrain~degrees:~}&~~~\Omega_L[\{\bk\}|\bx]&= \frac{1}{N}\log \sum_{\bc\in\allgraphs}W(\bc|\bx)\prod_i\delta_{\bk_i,\bk_i(\bc)}
\label{eq:Omega_k}
\\
\hspace*{-2mm}
{\rm constrain~statistics:}&~~~\Omega_L[P|\bx]&= \frac{1}{N}\log \sum_{\bc\in\allgraphs}W(\bc|\bx)\prod_{\bk,x}\delta\Big[P(\bk|x)-P(\bk|x,\bc)\Big]
\label{eq:Omega_Pk}
\\
\hspace*{-2mm}
&&
\hspace{-5mm}= \frac{1}{N}\log\!\sum_{\bk_1\ldots \bk_N} \prod_{\bk,x}\delta\Big[P(\bk|x)-\frac{\sum_i \delta_{\bk,\bk_i}\delta[x\!-\!x_i]}
{Np(x)}\Big]e^{N\Omega_L[\{\bk\}|\bx]}
\nonumber
\end{eqnarray}
The larger $\Omega_L[\ldots]$, the larger the effective number of graphs with the
imposed global topology, viz. $\{\bk_i\}$ or $P(\bk|x)$, so the less
specific is the proposed macroscopic topology characterization.
We will find that generally $\Omega_L[\ldots]=\order(N^0)$ as $N\to\infty$.
The remainder of this paper deals with the calculation of (\ref{eq:Omega_k}) and (\ref{eq:Omega_Pk}) in the limit $N\to\infty$, and
their dependence on the choices made for $P(\bk|x)$ and the for ensemble characteristics as defined by $p(x)$ and $Q(x,x^\prime)$.

\section{Asymptotic values of the entropy of network topology in a given ensemble}

\subsection{Derivation of steepest descent extremization formulas}

Since the ensemble (\ref{eq:ensemble}) is invariant under all node permutations, the difference between the two formulae (\ref{eq:Omega_k},\ref{eq:Omega_Pk}) should reflect only the node permutation freedom that is present in (\ref{eq:Omega_Pk}) but absent from (\ref{eq:Omega_k}). We evaluate (\ref{eq:Omega_k},\ref{eq:Omega_Pk}) by writing each Kronecker $\delta$ and each $\delta$-function in integral form.
Upon defining the short-hands $k_i^0=1$ for all $i$, and $\bomega_i\cdot\bk_i=\sum_{\ell=1}^{L}\omega_i^\ell k_i^\ell$,
expression (\ref{eq:generalized_degrees2}) allows us to simplify the term $\delta_{\bk,\bk_i(\bc)}$ to
\begin{eqnarray}
\delta_{\bk_i,\bk_i(\bc)}&=&\int_{-\pi}^\pi\!\frac{d\bomega_i e^{i\bomega_i\cdot\bk_i}}{(2\pi)^L}~e^{-i\sum_jc_{ij}\sum_{\ell=1}^L
\omega_i^\ell k_j^{\ell-1}}
\end{eqnarray}
We next define
\begin{eqnarray}
D[\{\bomega,\bk\}|\bx]&=&
\sum_{\bc\in\allgraphs}W(\bc|\bx)e^{-i\sum_{ij}c_{ij}\sum_{\ell=1}^L
\omega_i^\ell k_j^{\ell-1}}
\label{eq:D}
\end{eqnarray}
and subsequently find our that our two entropies can be written in the form
\begin{eqnarray}
\Omega_L[\{\bk\}|\bx]&=& \frac{1}{N}\log \int_{-\pi}^\pi\!\prod_i\Big[\frac{d\bomega_i e^{i\bomega_i\cdot\bk_i}}{(2\pi)^L}\Big]D[\{\bomega,\bk\}|\bx]
\label{eq:Omega_k2}
\\
\Omega_L[P|\bx]&=& \frac{1}{N}\log \int\!\prod_{\bk,x}\Big[\frac{d \hat{P}(\bk|x)e^{i N\hat{P}(\bk|x)P(\bk|x)}}{2\pi/N}\Big]
\sum_{\bk_1\ldots \bk_N} e^{N\Omega_L[\{\bk\}|\bx]-i\sum_i\hat{P}(\bk_i|x_i)}
 \nonumber
 \\
 &=& \frac{1}{N}\lim_{\Delta\to 0}\log
 \int\!\prod_{\bk,x}\Big[\frac{d \hat{P}(\bk|x)e^{i N\Delta\hat{P}(\bk|x)P(\bk|x)}}{2\pi/N\Delta}\Big]
 \nonumber
 \\
 &&\times
\sum_{\bk_1\ldots\bk_N} \int_{-\pi}^\pi\!\prod_i\Big[\frac{d\bomega_i e^{i[\bomega_i\cdot\bk_i-\hat{P}(\bk_i|x_i)/p(x_i)]}}{(2\pi)^L}\Big]
D[\{\bomega,\bk\}|\bx]
\label{eq:Omega_Pk2}
\end{eqnarray}
The core of the problem is apparently to calculate the function $D[\{\bomega,\bk\}]$ in (\ref{eq:D}), which involves the introduction of a measure
$W(\bomega,\bk,x|\{\bomega,\bk\})=N^{-1}\sum_i \delta_{\bk,\bk_i}\delta[x-x_i]\delta[\bomega-\bomega_i]$:
\begin{eqnarray}
D[\{\bomega,\bk\}|\bx]&=&\prod_{i<j}\left\{
1+\frac{c}{N}Q(x_i,x_j)\Big[e^{-i\sum_{\ell=1}^L[
\omega_i^\ell k_j^{\ell-1}+\omega_j^\ell k_i^{\ell-1}]}-1\Big]
\right\}
\nonumber
\\
&=& \exp\Big\{
\frac{1}{2}cN \int\!d xd x^\prime
Q(x,x^\prime)\int_{-\pi}^\pi\!d\bomega d\bomega^\prime \sum_{\bk\bk^\prime}
W(\bomega,\bk,x|\ldots)
\nonumber
\\
&&
\times W(\bomega^\prime\!,\bk^\prime\!,x^\prime|\ldots)
\Big[e^{-i\sum_{\ell=1}^L[
\omega_\ell k^\prime_{\ell-1}+\omega^\prime_\ell k_{\ell-1}]}-1\Big]+\order(N^0)\Big\}
\end{eqnarray}
We isolate $W(\ldots|\ldots)$ via suitable integrations
over $\delta$-functions, using  the functional measure $\{d W\}=\lim_{\Delta\bomega\to\bnull}\lim_{\Delta x\to 0}\prod_{\bomega,\bk,x}[d W(\bomega,\bk,x)\Delta\bomega\Delta x\sqrt{N/2\pi}]$, resulting in
\begin{eqnarray}
\hspace*{-15mm}
D[\{\bomega,\bk\}]
&=&\int\!\{d W d \hat{W}\} e^{i N\int_{-\pi}^\pi\!d\bomega d x\sum_{\bk} \hat{W}(\bomega,\bk,x)W(\bomega,\bk,x)+\order(N^0)}
\nonumber
\\
\hspace*{-15mm}
&&\times e^{\frac{1}{2}cN \int\!d xd x^\prime
  Q(x,x^\prime)\int_{-\pi}^\pi\!d\bomega d\bomega^\prime \sum_{\bk\bk^\prime}
W(\bomega,\bk,x)
 W(\bomega^\prime\!,\bk^\prime\!,x^\prime)
\Big[e^{-i\sum_{\ell=1}^L[
\omega_\ell k^\prime_{\ell-1}+\omega^\prime_\ell k_{\ell-1}]}-1\Big]}
\nonumber
\\
\hspace*{-15mm}
&&\times e^{-i\sum_i\hat{W}(\bomega_i,\bk_i,x_i)}
\label{eq:D_calculated}
\end{eqnarray}
Now only the last line contains microscopic variables, and it factorizes fully over the nodes of the network.  Upon inserting (\ref{eq:D_calculated})
into (\ref{eq:Omega_k2}) and (\ref{eq:Omega_Pk2}) this allows us to evaluate both expressions for $N\to\infty$ via steepest descent integration
over the distributions $W(\bomega,\bk,x)$, leading to
\begin{eqnarray}
\Omega_L[\{\bk\}|\bx]&=& {\rm extr}_{\{W,\hat{W}\}} \Psi_{1}[\{W,\hat{W}\}]
\\
\Omega_L[P|\bx]&=&
{\rm extr}_{\{W,\hat{W},\hat{P}\}} \Psi_{2}[\{W,\hat{W},\hat{P}\}]
\end{eqnarray}
with the functions
\begin{eqnarray}
\hspace*{-10mm}
\Psi_{1}[\{W,\hat{W}\}]&=&
i\int_{-\pi}^\pi\!d\bomega d x\sum_{\bk} \hat{W}(\bomega,\bk,x)W(\bomega,\bk,x)+\Phi[\{W\}]
\nonumber
\\
\hspace*{-10mm}
&& + \int\!d x~p(x)\sum_{\bk}P(\bk|x)\log
\int_{-\pi}^\pi\!\frac{d\bomega}{(2\pi)^L} ~e^{i[\bomega\cdot\bk-\hat{W}(\bomega,\bk,x)]}
\\
\hspace*{-10mm}
\Psi_{2}[\{W,\hat{W},\hat{P}\}]&=&
i\int_{-\pi}^\pi\!d\bomega d x\sum_{\bk} \hat{W}(\bomega,\bk,x)W(\bomega,\bk,x)+\Phi[\{W\}]\nonumber
\\
\hspace*{-10mm}&&+ i \int\!d x\sum_{\bk}\hat{P}(\bk|x)P(\bk|x)
\nonumber
\\
\hspace*{-10mm}
&&
 +\int\!d x~p(x)\log
\int_{-\pi}^\pi \frac{d\bomega}{(2\pi)^L} \sum_{\bk}e^{i[\bomega\cdot\bk-\hat{P}(\bk|x)/p(x)-\hat{W}(\bomega,\bk,x)]}
\end{eqnarray}
where
\begin{eqnarray}
\Phi[\{W\}]&=&
\frac{1}{2}c\int\!d xd x^\prime~ Q(x,x^\prime)\int_{-\pi}^\pi\!d\bomega
d\bomega^\prime \sum_{\bk\bk^\prime}
W(\bomega,\bk,x)
 W(\bomega^\prime\!,\bk^\prime\!,x^\prime)
 \nonumber
 \\
 &&\hspace*{30mm}\times
\Big[e^{-i\sum_{\ell=1}^L[
\omega_\ell k^\prime_{\ell-1}+\omega^\prime_\ell k_{\ell-1}]}-1\Big]
\label{eq:PhiW}
\end{eqnarray}
It will be convenient to introduce new functions $Q(\bk|x)=\exp[-i \hat{P}(\bk|x)/p(x)]$ and $V(\bomega,\bk,x)=\exp[-i \hat{W}(\bomega,\bk,x)]$ so that our saddle-point equations simplify to
\begin{eqnarray}
\Omega_L[\{\bk\}|\bx]&=& {\rm extr}_{\{V,W\}} \tilde{\Psi}_{1}[\{V,W\}]
\label{eq:Omega_k_in_Psi}
\\
\Omega_L[P|\bx]&=&
{\rm extr}_{\{Q,V,W\}} \tilde{\Psi}_{2}[\{Q,V,W\}]
\label{eq:Omega_Pk_in_Psi}
\end{eqnarray}
with the functions
\begin{eqnarray}
\hspace*{-15mm}
\tilde{\Psi}_{1}[\{V,W\}]&=&
\Phi[\{W\}]-\int_{-\pi}^\pi\!d\bomega d x\sum_{\bk} W(\bomega,\bk,x)\log V(\bomega,\bk,x)
\nonumber
\\
\hspace*{-15mm}
&& + \int\!d x~p(x)\sum_{\bk}P(\bk|x)\log
\int_{-\pi}^\pi\!\frac{d\bomega}{(2\pi)^L} ~V(\bomega,\bk,x)e^{i\bomega\cdot\bk}
\label{eq:final_Psi1}
\\
\hspace*{-15mm}
\tilde{\Psi}_{2}[\{Q,V,W\}]&=&
\Phi[\{W\}]-\int_{-\pi}^\pi\!d\bomega d x\sum_{\bk} W(\bomega,\bk,x)\log V(\bomega,\bk,x)\nonumber
\\
&&- \int\!d x~p(x)\sum_{\bk}P(\bk|x)\log Q(\bk|x)
\nonumber
\\
\hspace*{-15mm}
&&
 +\int\!d x~p(x)\log\sum_{\bk}Q(\bk|x)
\int_{-\pi}^\pi \frac{d\bomega}{(2\pi)^L}  V(\bomega,\bk,x)e^{i\bomega\cdot\bk}
\label{eq:final_Psi2}
\end{eqnarray}

\subsection{Simplification and reduction of the functional saddle-point equations}

We can now do the functional variations of $\tilde{\Psi}_{1}[\ldots]$ and $\tilde{\Psi}_{2}[\ldots]$ and find our saddle-point equations from which to solve $\{Q,V,W\}$. For $\tilde{\Psi}_1[\ldots]$ (referring to ensembles with constrained generalized degrees) these are found to be the following:
\begin{eqnarray}
\hspace*{-10mm}
\log V(\bomega,\bk,x)&=&
c\int\!d x^\prime Q(x,x^\prime)\int_{-\pi}^\pi\!d\bomega^\prime \sum_{\bk^\prime}
 W(\bomega^\prime\!,\bk^\prime\!,x^\prime)
\Big[e^{-i\sum_{\ell=1}^L[
\omega_\ell k^\prime_{\ell-1}+\omega^\prime_\ell k_{\ell-1}]}-1\Big]
\\
\hspace*{-10mm}
W(\bomega,\bk,x)
 &=& \frac{p(x)P(\bk|x)~V(\bomega,\bk,x)e^{i\bomega\cdot\bk}}{
\int_{-\pi}^\pi\!d\bomega^\prime~V(\bomega^\prime,\bk,x)e^{i\bomega^\prime\cdot\bk}}
\label{eq:W_solution}
\end{eqnarray}
For $\tilde{\Psi}_2[\ldots]$ (referring to ensembles with constrained distributions of generalized degrees) these are found to be the following:
\begin{eqnarray}
\hspace*{-10mm}
\log V(\bomega,\bk,x) &=&
c\int\!d x^\prime Q(x,x^\prime)\int_{-\pi}^\pi\!d\bomega^\prime \sum_{\bk^\prime}
 W(\bomega^\prime\!,\bk^\prime\!,x^\prime)
\Big[e^{-i\sum_{\ell=1}^L[
\omega_\ell k^\prime_{\ell-1}+\omega^\prime_\ell k_{\ell-1}]}-1\Big]
\\
\hspace*{-10mm}
W(\bomega,\bk,x)
 &=&
 \frac{p(x)Q(\bk|x)
 V(\bomega,\bk,x)e^{i\bomega\cdot\bk}}
 {\sum_{\bk^\prime}Q(\bk^\prime|x)
\int_{-\pi}^\pi \!d\bomega^\prime~ V(\bomega^\prime,\bk^\prime,x)e^{i\bomega^\prime\cdot\bk^\prime}}
\\
\hspace*{-10mm}
P(\bk|x)
&=&
\frac{Q(\bk|x)\int_{-\pi}^\pi\! d\bomega~
 V(\bomega,\bk,x)e^{i\bomega\cdot\bk}}
{\sum_{\bk^\prime}Q(\bk^\prime|x)
\int_{-\pi}^\pi\! d\bomega~  V(\bomega,\bk^\prime,x)e^{i\bomega\cdot\bk^\prime}}
\label{eq:PinQ}
\end{eqnarray}
The last equation is easily solved, viz.
\begin{eqnarray}
Q(\bk|x)&=&
\frac{P(\bk|x)}{\int_{-\pi}^\pi\! d\bomega~
 V(\bomega,\bk,x)e^{i\bomega\cdot\bk}}
 \label{eq:Q_found}
\end{eqnarray}
whereas in both cases (constrained degrees versus constrained degree statistics) we can eliminate immediately the kernels $W(\bomega,\bk,x)$, leaving us in either case with a closed problem for the kernel $V(\bomega,\bk,x)$ only. Upon inserting the solution (\ref{eq:Q_found}) into (\ref{eq:PinQ})
one finds that this remaining problem is in fact {\em identical} for both types of constraints, namely
\begin{eqnarray}
\log V(\bomega,\bk,x) &=&
c\int\!d x^\prime p(x^\prime) Q(x,x^\prime)\sum_{\bk^\prime}P(\bk^\prime|x^\prime)\nonumber
\\
&&\times
 \frac{\int_{-\pi}^\pi\!d\bomega^\prime~ V(\bomega^\prime,\bk^\prime,x^\prime)e^{i\bomega^\prime\cdot\bk^\prime}\Big[e^{-i\sum_{\ell=1}^L[
\omega_\ell k^\prime_{\ell-1}+\omega^\prime_\ell k_{\ell-1}]}-1\Big]}{
\int_{-\pi}^\pi\!d\bomega^\prime~V(\bomega^\prime,\bk^\prime,x^\prime)e^{i\bomega^\prime\cdot\bk^\prime}}
\label{eq:closed_for_V}
\end{eqnarray}
In addition one finds that (\ref{eq:W_solution}) holds in both cases.
The solution of (\ref{eq:closed_for_V}) is of the form
\begin{eqnarray}
V(\bomega,\bk,x)&=& e^{-c\int\!dx^\prime Q(x,x^\prime)p(x^\prime)}\exp\Big[c\sum_{\bxi\in\N^L}\gamma(\bk,\bxi,x) e^{-i\bomega\cdot\bxi}\Big]
\label{eq:V_in_gamma}
\end{eqnarray}
where $\gamma(\bk,\bxi,x)$ then obeys
\begin{eqnarray}
\gamma(\bk,\bxi,x) &=&
\int\!d x^\prime p(x^\prime) Q(x,x^\prime)\sum_{\bk^\prime}P(\bk^\prime|x^\prime)\prod_{\ell=1}^L \delta_{\xi_\ell,k^\prime_{\ell-1}}
\\
&&\times
 \frac{\int_{-\pi}^\pi\!d\bomega~\exp\Big[i\sum_{\ell=1}^L\omega_\ell( k_\ell^\prime- k_{\ell-1})+c\sum_{\bxi^\prime\in\N^L}\gamma(\bk^\prime,\bxi^\prime,x^\prime) e^{-i\bomega\cdot\bxi^\prime}\Big]}{
\int_{-\pi}^\pi\!d\bomega~\exp\Big[i\bomega\cdot\bk^\prime+c\sum_{\bxi^\prime\in\N^L}\gamma(\bk^\prime,\bxi^\prime,x^\prime) e^{-i\bomega\cdot\bxi^\prime}\Big]}
\nonumber
\end{eqnarray}
The two integrals over $\bomega$ in the latter fraction can be done. Both are of the form
\begin{eqnarray}
\hspace*{-10mm}
{\mathcal I}(\bk,\bk^\prime,x)&=&
\int_{-\pi}^\pi\!d\bomega~\exp\Big[i\bomega\cdot\bk+c\sum_{\bxi^\prime\in\N^L}\gamma(\bk^\prime,\bxi^\prime,x^\prime) e^{-i\bomega\cdot\bxi^\prime}\Big]
\nonumber\\
\hspace*{-10mm}
&=&(2\pi)^L\sum_{m\geq 0}
\frac{c^m}{m!}\sum_{\bxi^1\ldots\bxi^m \in\N^L}\Big[\prod_{n=1}^m\gamma(\bk^\prime,\bxi^n,x^\prime)\Big]  \delta_{\bk,\sum_{n\leq m}\bxi^n}
\end{eqnarray}
and hence the equation for $\gamma(\bk,\bxi,x)$ becomes
\begin{eqnarray}
\hspace*{-10mm}
\gamma(\bk,\bxi,x) &=& \delta_{\xi_1,1}
\int\!d x^\prime p(x^\prime) Q(x,x^\prime)\sum_{\bk^\prime}P(\bk^\prime|x^\prime)\prod_{\ell=1}^{L-1} \delta_{\xi_{\ell+1},k^\prime_{\ell}}
\\
\hspace*{-10mm}
&&
\times
 \frac{\sum_{m\geq 0}
\frac{c^m}{m!}\sum_{\bxi^1\ldots\bxi^m \in\N^L}\Big[\prod_{n=1}^m\gamma(\bk^\prime,\bxi^n,x^\prime)\Big]
\prod_{\ell=1}^L\delta_{k^\prime_\ell,k_{\ell-1}+\sum_{n\leq m}\xi_\ell^n}}{
\sum_{m\geq 0}
\frac{c^m}{m!}\sum_{\bxi^1\ldots\bxi^m \in\N^L}\Big[\prod_{n=1}^m\gamma(\bk^\prime,\bxi^n,x^\prime)\Big]  \delta_{\bk^\prime,\sum_{n\leq m}\bxi^n}}
\nonumber\\
\hspace*{-10mm}
&=& \delta_{\xi_1,1}
\int\!d x^\prime p(x^\prime) Q(x,x^\prime)\sum_{\bk^\prime}\frac{k_1^\prime}{c} P(\bk^\prime|x^\prime)\prod_{\ell=1}^{L-1} \delta_{\xi_{\ell+1},k^\prime_{\ell}}\nonumber
\\
\hspace*{-10mm}
&&
\times
 \frac{
\sum_{\bxi^1\ldots\bxi^{k_1^\prime-1} }\Big[\prod_{n=1}^{k_1^\prime-1}\gamma(\bk^\prime,\bxi^n,x^\prime)\Big]
\prod_{\ell=1}^L\delta_{k^\prime_\ell,k_{\ell-1}+\sum_{n< k_1^\prime}\xi_\ell^n}}{
\sum_{\bxi^1\ldots\bxi^{k_1^\prime} }\Big[\prod_{n=1}^{k_1^\prime}\gamma(\bk^\prime,\bxi^n,x^\prime)\Big]  \delta_{\bk^\prime,\sum_{n\leq k_1^\prime}\bxi^n}}
\end{eqnarray}
where we use the conventions that $[\prod_{n=1}^m u_n]_{m=0}\equiv 1$, $[\sum_{n=1}^m u_n]_{m=0}\equiv 0$, and $[\sum_{\xi_1\ldots\xi_m}u(\xi_1,\ldots,\xi_m)]_{m=0}\equiv 1$.
If $L=1$ we have $\bk=\to k$ and $\bxi\to 1$, so $\gamma(\bk,\bxi,x)\to \gamma(k,x)$. This describes the situation where the degrees
 are not generalized, but measure as usual only the number of direct links per node. Here our equation for $\gamma(\ldots)$ simplifies drastically to
\begin{eqnarray}
L=1:&~~~~&\gamma(k,x) =
\int\!d x^\prime p(x^\prime) Q(x,x^\prime)\sum_{k^\prime}k^\prime  \frac{P(k^\prime|x^\prime)}{
c\gamma(k^\prime,x^\prime)}
\label{eq:L1_gamma1}
\end{eqnarray}
The right-hand side is clearly independent of $k$, so $\gamma(k,x)=\gamma(x)$ with
\begin{eqnarray}
\gamma(x) &=&
\int\!d x^\prime p(x^\prime) \frac{Q(x,x^\prime)}{c\gamma(x^\prime)}\sum_{k}k P(k|x^\prime)
\label{eq:L1_gamma2}
\end{eqnarray}
If $L>1$ we can manipulate at most some further Kronecker $\delta$s, and
the final form is therefore
\begin{eqnarray}
\hspace*{-0mm}
\gamma(\bk,\bxi,x)
&=&  \frac{\xi_2 \delta_{\xi_1,1}}{c}
\int\!d x^\prime p(x^\prime) Q(x,x^\prime)\sum_{k^\prime\geq 0} P(\xi_2,\ldots,\xi_L,k^\prime|x^\prime)
\label{eq:final_eqn_for_gamma}
\\
\hspace*{-0mm}
&&\hspace*{-17mm}
\times
 \frac{
\sum_{\bxi^1\ldots\bxi^{\xi_2-1} }\Big[\prod_{n=1}^{\xi_2-1}\gamma((\xi_2,\ldots,\xi_L,k^\prime),\bxi^n,x^\prime)\Big]
\delta_{k^\prime,k_{L-1}+\sum_{n< \xi_2}\xi_L^n}
\prod_{\ell=1}^{L-1}\delta_{\xi_{\ell+1}-k_{\ell-1},\sum_{n<\xi_2}\xi_\ell^n}
}{
\sum_{\bxi^1\ldots\bxi^{\xi_2} }\Big[\prod_{n=1}^{\xi_2}\gamma((\xi_2,\ldots,\xi_L,k^\prime),\bxi^n,x^\prime)\Big]
\delta_{k^\prime,\sum_{n\leq \xi_2}\xi_L^n}
\prod_{\ell=1}^{L-1}\delta_{\xi_{\ell+1},\sum_{n\leq \xi_2}\xi_\ell^n}}
\nonumber
\end{eqnarray}

\subsection{Simplification of the asymptotic entropy formulas}

At this stage we insert our previous results for the kernels $\{V,W,Q\}$ into (\ref{eq:Omega_k_in_Psi},\ref{eq:Omega_Pk_in_Psi},\ref{eq:final_Psi1},\ref{eq:final_Psi2}) to arrive at more explicit expressions for the asymptotic entropies, which will only involve the function $\gamma(\bk,\bxi,x)$  of (\ref{eq:final_eqn_for_gamma}). The first step is to substitute
expression (\ref{eq:Q_found}) into (\ref{eq:final_Psi2}). This leads,
in combination with the fact that at the relevant saddle-points the
kernels $\{V,W\}$ obey identical equations for the two cases
(constrained generalized degrees versus constrained statistics of
generalized degrees), to the simple and natural   relation between our two entropies:
\begin{eqnarray}
\lim_{N\to\infty}\Omega_L[P|\bx]&=& \lim_{N\to\infty}\Omega_L[\{\bk\}|\bx]-\int\!d x ~p(x) \sum_{\bk}P(\bk|x)\log P(\bk|x)
\label{eq:entropy_relation}
\end{eqnarray}
The extra freedom to construct microscopic network realizations
in the case where we only constrain the generalized degree distribution, as opposed to
constraining the actual values of the generalized degrees, is measured
by the Shannon entropy of the imposed distribution $P(\bk|x)$.

The relation $(\ref{eq:entropy_relation})$ could be also derived from
the definition of $\Omega[\ldots]$, given in (\ref{eq:Omega_k})-(\ref{eq:Omega_Pk}).
In fact we can  observe that the probability $W(\bc|\bx)$
 present in the definition
(\ref{eq:Omega_k}) of $\Omega[\{\bk\}|\bx]$
 is invariant under all
permutations of the labels of those nodes that have the same ``hidden variable''
$x$; this follows directly  from definition (\ref{eq:ensemble}). Consequently $\Omega_L[\{\bk\}|\bx]$ must also be invariant
under any permutation of the labels of nodes with same value of
$x$.
It follows that $\Omega_L[\{\bk\}|\bx]$ is  dependent
on the degree sequence $\{\bk\}$ only through  the distributions  $\{P(\bk|x)\}$.
Therefore, we can use this simple insight  to  predict the relation between
$\Omega_L[\{\bk\}|\bx]$ and $\Omega_L[P|\bx]$ $(\ref{eq:entropy_relation})$.
In fact, because $\Omega_L[\{\bk\}|\bx]$ must be only dependent on
the distribution  $P(\bk|x)$, we
have that
\begin{eqnarray}
\Omega_L[P|\bx]&=& \frac{1}{N}\log\!\sum_{\bk_1\ldots \bk_N} \prod_{\bk,x}\delta\Big[P(\bk|x)-\frac{\sum_i \delta_{\bk,\bk_i}\delta[x\!-\!x_i]}
{Np(x)}\Big]e^{N\Omega_L[\{\bk\}|\bx]}
\nonumber\\
& =&\frac{1}{N} \log
e^{N\Omega_L[\{\bk\}|\bx]}\prod_x\frac{[Np(x)]!}{\prod_{\bf k}[
  Np(x)P(\bk|x)]!}.
\label{eq1}
\end{eqnarray}
where $\{\bk\}$ is any generalized degree sequence with degree distributions $P(\bk|x)$.
Using $(\ref{eq1})$ we can derive relation $(\ref{eq:entropy_relation})$.

In order to evaluate $\lim_{N\to\infty}\Omega_L[\{\bk\}|\bx]$ we  only need to express
$\lim_{N\to\infty}\Omega_L[\{\bk\}|\bx]$ in terms of the function $\gamma(\bk,\bxi,x)$. We first note that at the relevant saddle-point
the function $\Phi[\{W\}]$ (\ref{eq:PhiW}) takes the value
\begin{eqnarray}
\Phi[\{W\}]&=&
\frac{1}{2}\int\!d x\int_{-\pi}^\pi\!d\bomega \sum_{\bk}
W(\bomega,\bk,x)\log V(\bomega,\bk,x)
\end{eqnarray}
Insertion into (\ref{eq:final_Psi1}), followed by elimination of $W(\bomega,\bk,x)$ via (\ref{eq:W_solution}), leads us to
\begin{eqnarray}
\hspace*{-15mm}
\lim_{N\to\infty}\Omega_L[\{\bk\}|\bx]&=&
\int\!d x~p(x)\sum_{\bk}P(\bk|x)\log
\int_{-\pi}^\pi\!\frac{d\bomega}{(2\pi)^L} ~V(\bomega,\bk,x)e^{i\bomega\cdot\bk}
\nonumber
\\
\hspace*{-15mm}
&&
-\frac{1}{2}\int\!d x~ p(x)\sum_{\bk}P(\bk|x)
\frac{\int_{-\pi}^\pi\!d\bomega~e^{i\bomega\cdot\bk} V_0(\bomega,\bk,x)\log V(\bomega,\bk,x)}{
\int_{-\pi}^\pi\!d\bomega~V_0(\bomega,\bk,x)e^{i\bomega\cdot\bk}}
\end{eqnarray}
where $V_0(\bomega,\bk,x)=V(\bomega,\bk,x)\exp[c\int\!dx^\prime Q(x,x^\prime)p(x^\prime)]$.
The final step is the elimination of $V(\bomega,\bk,x)$ via (\ref{eq:V_in_gamma}), followed by integration over $\bomega$, using
the property that $\gamma(\bk,\bxi,x)=0$ unless $\xi_1=1$:
\begin{eqnarray}
\hspace*{-2mm}
~~~~~~~~~~~ \int_{-\pi}^\pi\!\frac{d\bomega}{(2\pi)^L} ~V_0(\bomega,\bk,x)e^{i\bomega\cdot\bk}&=&
\sum_{m\geq 0}\frac{c^m}{m!}\sum_{\bxi^1\ldots\bxi^m} \prod_{n\leq m}\!\Big[\gamma(\bk,\bxi^n\!,x)\Big]\delta_{\bk,\sum_{n\leq m}\bxi^n}
\nonumber
\\
\hspace*{-2mm}
&&\hspace*{-20mm}=\delta_{\bk,\bnull}+ \frac{c^{k_1}\theta[k_1\!-\!\frac{1}{2}]}{k_1!}\!\!\sum_{\bxi^1\ldots\bxi^{k_1}} \prod_{n\leq k_1}\!\Big[\gamma(\bk,\bxi^n\!,x)\Big]\delta_{\bk,\sum_{n\leq k_1}\bxi^{n}}
\\
\hspace*{-2mm}
 \int_{-\pi}^\pi\!\frac{d\bomega}{(2\pi)^L} ~V_0(\bomega,\bk,x)\log V_0(\bomega,\bk,x)e^{i\bomega\cdot\bk}&=&
\sum_{m>0}\frac{c^m}{(m\!-\!1)!}\sum_{\bxi^1\ldots\bxi^{m}} \prod_{n\leq m}\!\Big[\gamma(\bk,\bxi^n\!,x)\Big]\delta_{\bk,\sum_{n\leq m}\bxi^n}
\nonumber
\\
\hspace*{-2mm}
&&\hspace*{-20mm}= \frac{c^{k_1}\theta[k_1\!-\!\frac{1}{2}]}{(k_1\!-\!1)!}\!\sum_{\bxi^1\ldots\bxi^{k_1}} \prod_{n\leq k_1}\!\Big[\gamma(\bk,\bxi^n\!,x)\Big]\delta_{\bk,\sum_{n\leq k_1}\bxi^{n}}
\end{eqnarray}
So one arrives at  the compact result, where we have used the fact that if $k_1=0$ then $k_\ell=0$ for all $\ell$ (which follows from the definition
of the generalized degrees):
\begin{eqnarray}
\lim_{N\to\infty}\Omega_L[\{\bk\}|\bx]&=&
\sum_{k_1}P(k_1)\log\pi_c(k_1)+\frac{1}{2}\Big[c-\int\!dx~p(x)\sum_{\bk}k_1P(\bk|x)\Big]
\label{eq:FinalOmegaL}
\\
&&
\hspace*{-15mm}
+
\int\!d x~p(x)\sum_{\bk}P(\bk|x)\log\Big\{
\sum_{\bxi^1\ldots\bxi^{k_1}}\! \Big[\prod_{n\leq k_1}\gamma(\bk,\bxi^n\!,x)\Big]\delta_{\bk,\sum_{n\leq k_1}\bxi^{n}}\Big\}
\nonumber
\end{eqnarray}
with the average-$c$ Poissonian degree distribution $\pi_c(k)=c^k e^{-c}/k!$.

\section{Applications of the general theory}

\subsection{Regular random graphs}

Our first application domain is that of $r$-regular degree distribution $P(\bk|x)=\delta_{\bk,\bk(r)}$, with $\bk(r)=(r,r^2,\ldots,r^L)$.
 Here one can solve (\ref{eq:final_eqn_for_gamma}) explicitly:
\begin{eqnarray}
\hspace*{-2mm}
\gamma(\bk,\bxi,x)
&=&  \frac{\xi_2 \delta_{\xi_1,1}}{c}
\int\!d x^\prime p(x^\prime) Q(x,x^\prime)\sum_{k^\prime\geq 0} \delta_{(\xi_2,\ldots,\xi_L,k^\prime),\bk(r)}
\\
\hspace*{-2mm}
&&
\hspace*{-5mm}\times
 \frac{
\sum_{\bxi^1\ldots\bxi^{k_1(x^\prime)-1} }\Big[\prod_{n=1}^{r-1}\gamma(\bk(r),\bxi^n,x^\prime)\Big]
\delta_{k_L(r),k_{L-1}+\sum_{n< r}\xi_L^n}
\prod_{\ell=1}^{L-1}\delta_{r^\ell-k_{\ell-1},\sum_{n<r}\xi_\ell^n}
}{
\sum_{\bxi^1\ldots\bxi^{k_1(x^\prime)} }\Big[\prod_{n=1}^{r}\gamma(\bk(r),\bxi^n,x^\prime)\Big]
\prod_{\ell=1}^{L}\delta_{r^\ell,\sum_{n\leq r}\xi_\ell^n}}
\nonumber
\end{eqnarray}
The solution is seen to be of the form $\gamma(\bk,\bxi,x)=\gamma(\bk,x)\delta_{\bxi,(1,r,r^2,\ldots,r^{L-1})}$, and independent of $k_L$.
Insertion of this form into the above equation then gives
\begin{eqnarray}
\gamma(\bk,x)
&=&  \frac{r}{c}
\int\!d x^\prime p(x^\prime) Q(x,x^\prime)
 \frac{
\prod_{\ell=1}^{L-1}\delta_{k_{\ell},r^{\ell}}
}{\gamma(\bk(r),x^\prime)}
\end{eqnarray}
We conclude that
\be
 \gamma(\bk,\bxi,x)=\gamma(x)\Big[\prod_{\ell=1}^{L-1}\delta_{k_\ell,r^\ell}\Big]\Big[\prod_{\ell=1}^L\delta_{\xi_\ell,r^{\ell-1}}\Big]
\ee
where $\gamma(x)$ is the solution of
\begin{eqnarray}
\gamma(x)
&=&  \frac{r}{c}
\int\!d x^\prime p(x^\prime) \frac{Q(x,x^\prime)}{\gamma(x^\prime)}
\end{eqnarray}
For the entropies (\ref{eq:entropy_relation},\ref{eq:FinalOmegaL}) one then finds
\begin{eqnarray}
&&
\lim_{N\to\infty}\Omega_L[P|\bx]=\lim_{N\to\infty}\Omega_L[\{\bk\}|\bx]=
\log\pi_c(r)
+r\int\!d x~p(x)\log \gamma(x)+\frac{1}{2}(c-r)
\end{eqnarray}
As expected, the two entropies are identical (since for regular graphs there is no entropy contribution from degree permutations)
and independent of $L$ (since upon specifying that the degrees are $r$-regular, the full distributions $P(\bk|x)$ are uniquely specified
for any $L$).
In the special case $Q(x,x^\prime)=1$  of uncorrelated degrees the above solution simplifies further. Now $\gamma(x)=\sqrt{r/c}$, and
\begin{eqnarray}
\lim_{N\to\infty}\Omega_L[P|\bx|\bx]
=\lim_{N\to\infty}\Omega_L[\{\bk\}|\bx]=
\log\pi_c(r)
+\frac{1}{2}r\log (r/c)+\frac{1}{2}(c-r)
\end{eqnarray}

\subsection{The case $L=1$}

For $L=1$ we have already simplified our formula for the function $\gamma(\bk,\bxi,x)$ to relation (\ref{eq:L1_gamma2}) for a simple function
$\gamma(x)$.  We can do the same for expression (\ref{eq:FinalOmegaL}) for the entropy, which gives
\begin{eqnarray}
\hspace*{-12mm}
\lim_{N\to\infty}\Omega_1[\{k\}|\bx]&=&
\sum_{k}P(k)\log\pi_c(k)+\frac{1}{2}(c\!-\!\overline{k})
+
\int\!d x~p(x)\sum_{k} P(k|x)\log
\gamma^k(x)
\label{eq:L1_result}
\\
\hspace*{-12mm}
\lim_{N\to\infty}\Omega_1[P|\bx]&=& \sum_{k}P(k)\log\pi_c(k)+\frac{1}{2}(c\!-\!\overline{k})
-
\int\!d x~p(x)\sum_{k} P(k|x)\log[P(k|x)/\gamma^k(x)]
\label{eq:omega_L1}
\end{eqnarray}
with $\pi_c(k)=c^k e^{-c}/k!$, with $\overline{k}=\int\!dx~p(x)\sum_k
kP(k|x)$, $P(k)=\int dx p(x)P(k|x)$  and
 where $\gamma(x)$ is to be solved from
\begin{eqnarray}
\gamma(x) &=&
\int\!d x^\prime p(x^\prime) \frac{Q(x,x^\prime)}{c\gamma(x^\prime)}\sum_{k}k P(k|x^\prime)
\label{eq:gamma_L1}
\end{eqnarray}
We see immediately that for $Q(x,x^\prime)=1$ (the Erd\"{o}s-R\'{e}nyi ensemble), and upon choosing $P(k|x)=P(k)$ (since for $Q(x,x^\prime)=1$ the hidden variables $x$ are obsolete)  we would have had $\gamma(x)=\sqrt{\overline{k}/c}$ $\forall x$. Expression  (\ref{eq:L1_result}) now becomes
\begin{eqnarray}
\hspace*{-10mm}
Q(x,x^\prime)=1:&~&
\begin{array}{lll}
\lim_{N\to\infty}\Omega_1[\{k\}|\bx]&\!\!=\!\!&
\sum_{k}P(k)\log\pi_c(k)+\frac{1}{2}(c\!-\!\overline{k}) +\frac{1}{2}\overline{k}\log (\overline{k}/c)
\\[2mm]
\lim_{N\to\infty}\Omega_1[P|\bx]&\!\!=\!\!&
-\sum_{k}P(k)\log[P(k)/\pi_c(k)]+\frac{1}{2}(c\!-\!\overline{k})+\frac{1}{2}\overline{k}\log (\overline{k}/c)
\end{array}
\label{eq:Poissonnian}
\end{eqnarray}
So, if one also chooses $\overline{k}=c$, the entropy of networks with degree distribution $P(k)$ in the Erd\"{o}s-R\'{e}nyi ensemble is minus the Kullback-Leibler distance
between $P(k)$ and a Poisson degree distribution, provided the ensemble and $P(k)$ have the same average connectivity.
An alternative derivation of equation $(\ref{eq:Poissonnian})$ can also be obtained  starting from the
expression of the total number of graphs with given degree
sequence ${\cal N}[\{\bk\}]$ derived in \cite{Enumeration1,Wormald,europhys}:
\be
{\cal N}[\{\bk\}]=(\overline{k}N-1)!!\frac{e^{-\frac{1}{4}\lambda}}{\prod_i
  k_i!}
\label{eq:sigma}
\ee
with $\lambda=\left({\overline{k^2}}/{\overline{k}}\right)^2-1$
and $\overline{k^2}=\sum_i k_i^2/N$.
The entropy $\Omega[\{\bk\}]$ of the degree sequence $\{\bk\}$ in the
 Erd\"{o}s-R\'{e}nyi  ensemble, is the logarithm of the probability of
having  one of the total number ${\cal N}[\{k\}]$ of possible networks
in the ensemble.
Since in a  Erd\"{o}s-R\'{e}nyi network each link has a probability $c/N$ to be
present, we have
\begin{equation}
\lim_{N\rightarrow \infty}\Omega_1[\{\bk\}]=\lim_{N\rightarrow \infty}
\frac{1}{N}\log\left\{{\cal N}[\{k\}]\left(\frac{c}{N}\right)^{N\overline{k}/2}\left(1-\frac{c}{N}\right)^{N(N-1)/2-N\overline{k}/2}\right\}.
\label{eq:entropy_sigma}
\end{equation}
Upon inserting the expression of ${\cal N}[\{\bk\}]$,  $(\ref{eq:sigma})$
in Eq. $(\ref{eq:entropy_sigma})$  we recover $(\ref{eq:Poissonnian})$.

The other terms in (\ref{eq:L1_result}) apparently
represent the effect of average connectivity mismatches and of the degree correlations induced by $Q(.,.)$, and make matters more complicated.
The simple form of our $L=1$ equations, however, still allows us to push the analysis further for certain cases, by solving $\gamma(x)$
explicitly from equation (\ref{eq:gamma_L1}). For instance, if the (symmetric) kernel $Q(x,x^\prime)$ has an eigenfunction  $f(x)=\sqrt{p(x)k(x)}$, with $k(x)=\sum_k k P(k|x)$ then
 \begin{eqnarray}
 \int\!d x^\prime Q(x,x^\prime)f(x^\prime)=\lambda f(x),~~f(x)=\sqrt{p(x)k(x)}:&~~~&\gamma(x)=\sqrt{\frac{\lambda}{c}}f(x)
 \label{eq:Q_eigenfunctions}
 \end{eqnarray}
 If $Q(.,.)$ has this property, together with the normalization
 $\int\!d xd x^\prime p(x)Q(x,x^\prime)p(x^\prime)=1$, then one finds
 that the entropy  (\ref{eq:L1_result}) becomes
 \begin{eqnarray}
 \hspace*{-3mm}
\lim_{N\to\infty}\Omega_1[\{k\}]|\bx]&=&
\sum_{k}P(k)\log\pi_c(k)+\frac{1}{2}(c\!-\!\overline{k})
+
\frac{1}{2}\int\!d x~p(x)k(x)\log[\lambda p(x)k(x)/c]
\nonumber
\\
\hspace*{-3mm}
&=& \sum_{k}P(k)\log\pi_c(k)+\frac{1}{2}(c\!-\!\overline{k})
+
\frac{1}{2}\int\!d x~p(x)k(x)\log[p(x)k(x)]
+
\frac{1}{2}\overline{k}\log[\lambda /c]
\nonumber
\\
\hspace*{-3mm}&&
\label{eq:L1_result_specialQ}
\end{eqnarray}
(where $\overline{k}=\int\!d x~p(x)k(x)$).
Let us next discuss some example kernels $Q(x,x^\prime)$ for which $\gamma(x)$ can be solved explicitly, either directly, or via the above procedure based
on using eigenfunctions of $Q(.,.)$:
\begin{itemize}
\item
First example:\\
Here we assume $Q(x,x^\prime)$ to be such that the conditional connectivities $k(x)=\sum_k kP(k|x)$ are
the typical ones for the ensemble
(\ref{eq:ensemble}), which implies that
\begin{equation}
k(x)= c \int\! d x^\prime~ Q(x,x^\prime)p(x^\prime) \nonumber
\end{equation}
and $\overline{k}=c$.
In this case (\ref{eq:gamma_L1}) has the  solution
$\gamma(x)=k(x)/c$, which leads to the following simple expression for
the entropies:
\begin{eqnarray}
\lim_{N\to\infty}\Omega_1[\{k\}|\bx]&=&
\int\!d x~p(x)\sum_{k} P(k|x)\log
\pi_{k(x)}
\label{eq:Omega_k_example3}
\\
\lim_{N\to\infty}\Omega_1[P|\bx]&=& -\int\!d x~p(x)\sum_{k} P(k|x)\log[P(k|x)/\pi_{k(x)}]
\label{eq:Omega_Pk_example3}
\end{eqnarray}
This indicates that in this case the entropy
$\lim_{N\to\infty}\Omega_1[P|\bx]$ takes the form of an integral over
$p(x)$ of the Kullback-Leibler distance between the probabilities
$P(k|x)$ and the Possion distribution $\pi_{k(x)}$.
We note that for the
hidden variable model the typical degree
distribution of the nodes with hidden variable $x$ is indeed $\pi_{k(x)}$
\cite{Boguna}.
\item
Second example:
\begin{eqnarray}
Q(x,x^\prime)=a_0+a_1\delta(x-x^\prime),~~~~~~k(x)=\frac{1}{2}c/p(x)
\end{eqnarray}
with $x\in[-1,1]$.
Normalization of $Q(.,.)$ tells us that $a_0=1-a_1\int\!d x~ p^2(x)$, and we need $0\leq a_1\leq [\int\!d x~ p^2(x)]^{-1}$
to ensure non-negative bond probabilities in our network ensemble.
The networks in this ensemble have a non trivial community
structure. In fact  nodes with same hidden-variable have a larger
probability to be connected.
Here one finds a solution with $\overline{k}=c$ and
$\gamma(x)=\gamma$, where
\begin{eqnarray}
\hspace*{-5mm}&&
\gamma= \sqrt{1-a_1\int\!d x ~p^2(x)+\frac{1}{2}a_1}
\\
\hspace*{-5mm}&&
\lim_{N\to\infty}\Omega_1[\{k\}|\bx]=
\sum_{k}P(k)\log\pi_c(k)
+
\frac{1}{2}c\log\Big[1\!-\!a_1\!\int_{-1}^1\!d x ~p^2(x)\!+\!\frac{1}{2}a_1\Big]
\end{eqnarray}

\item
Third example:
\begin{eqnarray}
Q(x,x^\prime)&=& \frac{g(x)+g(x^\prime)}{2\int\!d x^\pprime p(x^\pprime)g(x^\pprime)},~~~~~~
k(x)=\frac{\Big[\sqrt{\bra g^2\ket_0}\!+\!g(x)\Big]^2}{p(x)}
\end{eqnarray}
with $x\in[-1,1]$, with the short-hand $\bra \phi\ket_0=\frac{1}{2}\int_{-1}^1\!d x~\phi(x)$, and with $g(x)\geq 0$ for all $x\in[-1,1]$.
Here one finds the solution
\begin{eqnarray}
\gamma(x)&=& \frac{1}{\sqrt{c}} \sqrt{\frac{\bra g\ket_0\!+\!\sqrt{\bra g^2\ket_0}}{\int\!d x^\prime~p(x^\prime)g(x^\prime)}}~ \Big[\sqrt{\bra g^2\ket_0}\!+\!g(x)\Big],~~~~~
\lambda=\frac{\bra g\ket_0\!+\!\sqrt{\bra g^2\ket_0}}{\int\!d x~p(x)g(x)}
\end{eqnarray}
\end{itemize}

\subsection{The case $L=2$}

Here we have to find first the solution of (\ref{eq:final_eqn_for_gamma}), which now reduces to
\begin{eqnarray}
\hspace*{-15mm}
\gamma((k_1,k_2),(1,\xi),x)
&=&  \frac{\xi}{c}
\int\!d x^\prime p(x^\prime) Q(x,x^\prime)\sum_{k^\prime\geq 0} P(\xi,k^\prime|x^\prime)
\nonumber
\\
\hspace*{-15mm}
&&
\times
 \frac{
\sum_{\xi_1\ldots\xi_{\xi-1} }\Big[\prod_{n=1}^{\xi-1}\gamma((\xi,k^\prime),(1,\xi_n),x^\prime)\Big]
\delta_{k^\prime,k_1+\sum_{n< \xi}\xi_n}}
{
\sum_{\xi_1\ldots\xi_{\xi} }\Big[\prod_{n=1}^{\xi}\gamma((\xi,k^\prime),(1,\xi_n),x^\prime)\Big]
\delta_{k^\prime,\sum_{n\leq \xi}\xi_n}}
\end{eqnarray}
We observe that the right-hand side is independent of $k_2$, so the solution of our equation must have the form
$\gamma((k_1,k_2),(1,\xi),x)=\gamma(k_1,\xi,x)$, where
\begin{eqnarray}
\gamma(k,\xi,x)
&=&  \frac{\xi}{c}
\int\!d x^\prime p(x^\prime) Q(x,x^\prime)\sum_{k^\prime\geq 0} P(\xi,k^\prime|x^\prime)
\nonumber
\\
&&
\times
 \frac{
\sum_{\xi_1\ldots\xi_{\xi-1} }\Big[\prod_{n=1}^{\xi-1}\gamma(\xi,\xi_n,x^\prime)\Big]
\delta_{k^\prime,k+\sum_{n< \xi}\xi_n}}
{
\sum_{\xi_1\ldots\xi_{\xi} }\Big[\prod_{n=1}^{\xi}\gamma(\xi,\xi_n,x^\prime)\Big]
\delta_{k^\prime,\sum_{n\leq \xi}\xi_n}}.
\end{eqnarray}
The entropy would become
\begin{eqnarray}
\hspace*{-15mm}
\lim_{N\to\infty}\Omega_2[\{\bk\}|\bx]&=&
\sum_{k_1}P(k_1)\log\pi_c(k_1) +\frac{1}{2}\Big[c-\int\!dx~p(x)\sum_{k_1k_2}k_1 P(k_1,k_2|x)\Big]
\nonumber
\\
&&
\hspace*{-15mm}
+
\int\!d x~p(x)\sum_{k_1 k_2}P(k_1,k_2|x)\log\Big\{
\sum_{\xi_1\ldots\xi_{k_1}}\! \Big[\prod_{n\leq k_1}\gamma(k_1,\xi_n,x)\Big]
\delta_{k_2,\sum_{n\leq k_1}\xi_{n}}\Big\}
\end{eqnarray}

Let us limit ourselves to the simplest scenario where there are no degree correlations, i.e. $Q(x,x^\prime)=1$. Here we have $\gamma(k,\xi,x)
 =\gamma(k,\xi)$, and we need only the generalized degree statistics $P(k_1,k_2)=\int\!d x~p(x)P(k_1,k_2|x)$. Our formulae thereby reduce to
\begin{eqnarray}
\gamma(k,\xi)
&=&  \frac{\xi}{c}
 \sum_{k^\prime\geq 0} P(\xi,k^\prime)
 \frac{
\sum_{\xi_1\ldots\xi_{\xi-1} }\Big[\prod_{n=1}^{\xi-1}\gamma(\xi,\xi_n)\Big]
\delta_{k^\prime,k+\sum_{n< \xi}\xi_n}}
{
\sum_{\xi_1\ldots\xi_{\xi} }\Big[\prod_{n=1}^{\xi}\gamma(\xi,\xi_n)\Big]
\delta_{k^\prime,\sum_{n\leq \xi}\xi_n}}
\\
\lim_{N\to\infty}\Omega_2[\{\bk\}|\bx]&=&
\sum_{k_1}P(k_1)\log\pi_c(k_1)
+\frac{1}{2}\Big[c-\sum_{k_1k_2}k_1 P(k_1,k_2)\Big]
\\
&&
\hspace*{-15mm}
+\sum_{k_1 k_2}P(k_1,k_2)\log\Big\{
\sum_{\xi_1\ldots\xi_{k_1}}\! \Big[\prod_{n\leq k_1}\gamma(k_1,\xi_n)\Big]
\delta_{k_2,\sum_{n\leq k_1}\xi_{n}}\Big\}
\nonumber
\end{eqnarray}
Here one observes the validity of the following simple relation:
\begin{eqnarray}
\sum_{k_2}\gamma(k_1,k_2)\gamma(k_2,k_1)&=&\frac{k_1}{c}P(k_1)
\end{eqnarray}

\section{Conclusions}

In conclusion, we have calculated the entropies
$\Omega_L[\{\bk\}|\bx]$ and $\Omega_L[P|\bx]$ of hierarchical constrained network topologies
in the ``canonical'' ensemble of large sparse networks described in terms of
"hidden variables".

The expression of the entropy $\Omega_L[P|\bx]$ assumes a very clear form
in the case in which the network topology under study is the degree
distribution of a
network of  the ensemble. Here the entropy  measures the large deviation of the topology of
the given networks from the typical topology of networks in the chosen
ensemble.

The entropy measures the likelihood that a particular network
topology belongs to an ensemble, as such it is an important quantity whenever one seeks to represent
or characterize observed networks in terms of appropriate random network ensembles.
We therefore believe that it may have many
applications in the future in the context of community detection problems as well as
other inference problems on complex networks.

\section*{Acknowledgements}

We thank our anonymous referees for suggesting useful alternative derivations in places, 
and for noting an error in the first version.
One of the authors (CJPV) acknowledges financial
support from project FIS2006-13321-C02-01) and grant PR2006-0458.
This work was also supported  by the project IST STREP GENNETEC contract No.034952.

\end{document}